\documentstyle[12pt]{article}
\begin{document}
\begin{center}
{\Large Pre-big bang model has Planck problem}\\
\bigskip
{\bf D.H. Coule}\\
\bigskip
School of Mathematical Sciences\\
University of Portsmouth, Mercantile House\\ 
Hampshire Terrace, Portsmouth PO1 2EG
\bigskip
\begin{abstract}
The pre-big bang's kinetic driven 
inflationary mechanism is not an adequate form of 
inflation: the Planck length grows more rapidly than the scale factor. 
In order to  
explain our ${\em large}$ universe,
the resulting post-big bang universe requires the 
same unnatural constants (Planck problem) 
as those of any other non-inflationary big bang model. 

\end{abstract}


\end{center}
\newpage
{\bf 1 Introduction}\\
The pre-big bang scenario, inspired by superstring theory, 
is an alternative inflationary universe
model to that of the usual scalar potential 
driven one - for reviews see [1]. This expansion can
now start at time $t\rightarrow -\infty$ and is 
of the form $a\sim (-t) ^p$ with $p<0$. 
This kinetic or pole-law inflation is a more  
general phenomena that occurs in various alternative gravity theories
such as induced gravity [2]. 

Most interest has developed in string theory where such an expansion
is driven by the dilaton's kinetic
energy; alternatively the model can be recast as a Brans-Dicke theory
with parameter $\omega=-1$.  

Most of the studies of this model have concentrated on the singularity
that occurs as $t\rightarrow  0_-$ and how one can undergo a branch
change to get to a less expansionary post-big bang region, now 
with expansion rate 
$a\sim t^{p}$ , $0<p\leq 1$. This branch change  has proven 
difficult to implement [3] but one hopes that  
additional terms in the string action could be responsible [4] 
or else appeal to
quantum cosmological notions such as tunneling [5]. We will not
be concerned with the actual mechanism of the 
branch change but assume it can be 
correctly accounted for. 

The simple model we consider is that from the low-energy string theory
with action [6]
\begin{equation} 
S=\int d^4x \sqrt{-g}\; \exp(-\phi) \left ( R-\omega  (\partial_{\mu}\phi)^2  
\right )\;\;.
\end{equation}
  We have only included the 
  dilaton $\phi$ term as this is the fundamental component that can possibly
  drive an expansion.  Using a field redefinition $\Phi=\exp(-\phi)$ the
  action can be rewritten in the more usual Brans-Dicke form
\begin{equation}
S=\int d^4x\sqrt{-g}\left ( \Phi R-\frac{\omega}{\Phi}(\partial _
{\mu}\Phi)^2\right )\;\;.
\end{equation}
For stability in Lorentzian space one requires $\omega>-3/2$. 
  Duality symmetry of string theory requires
  $\omega=-1$ [7] but we keep this $\omega$ term general for now. 

We will only be interested in  the  FRW ansatz
\begin{equation}
ds^2=-dt^2+a^2(t)d\Omega^2_3 \;,
\end{equation}
and will use Planck units throughout. We should stress that although Newton's
constant $G$ and its corresponding Planck length ($l_p\equiv G^{1/2}$) 
are time dependant 
in Brans-Dicke theory, we take Planck units to be their present values. 

The Brans-Dicke model is known to slow down the expansion rate of a 
potential driven inflationary model: the field $\Phi$ 
can gain kinetic energy
which drains energy from that being used for 
expansion -see eg.[8]. For $\omega<1/2$  
the conventional route of inflation becomes impossible to implement: any
scalar potential becomes too steep to have inflation [9]. If instead the  
kinetic energy is 
coupled to the gravitational field it can possibly drive the expansion. 
In the pre-big bang
scenario there are two types of solution starting at time
$t=-\infty$, one which starts at infinite size and collapses 
$a\sim (-t)^{1/\sqrt{3}}$ and another 
which has the required inflationary expansion $a\sim (-t)^{-1/\sqrt{3}}$. 

Brans-Dicke theory can be understood in either the Jordan frame or 
conformally transformed to alternative frames such as the so-called
Einstein frame [10]. In this Einstein 
frame it simply takes the form of a massless scalar
field which is not considered inflationary. We
will see that both frames are equivalent for understanding the pre-big bang
phase and that a deficiency of kinetic driven inflation is apparent in
either frame.  

{\bf 2 Conformal transformation to the Einstein frame} \\
By means of a conformal transformation of the form -see eg. [8,10]
\begin{equation}
\tilde{g}_{\mu\nu}=\Omega^2 g_{\mu\nu}
\end{equation}
where $\Omega^2=\Phi$ ,
we can find an equivalent action to expression (2), 
which can be expressed as
\begin{equation}
S=\int d^4x \sqrt{-\tilde{g}} \left ( R(\tilde{g}) -1/2 (\tilde{\nabla}
\sigma)^2\right )\;\;,
\end{equation}
where the scalar field $\sigma$ can be  defined from [8,10]
\begin{equation}
\Phi=\exp (\beta\sigma)\;\;,
\end{equation}
 and  $\beta^2=1/(2\omega+3)$.
 This action is simply that of a massless scalar field 
 whose field equations with a FRW metric are,
 \begin{equation}
 H^2+\frac{k}{a^2} = \dot{\sigma}^2
 \end{equation}
 \begin{equation}
 \ddot{\sigma} +3\frac{\dot{a}}{a} \dot{\sigma} =0\;\;\; \Rightarrow \;\;
 \dot{\sigma}^2 =\frac{A}{a^6}
 \end{equation}
 The solution of this equation ignoring the curvature is simply
 $a=A^{1/3} |t|^{1/3}$, with $t=0$ being the initial singularity.
 We can now highlight a serious problem that occurs with a big bang model
 that uses this matter source.
 If such a model is to account for our present universe then the
 constant $A$ has to be extremely large $\sim 10^{120}$ 
 in Planck units. This is true for
 any big bang model with a 
 matter source that obeys the strong energy condition 
 (eg. dust or radiation). For a perfect fluid with equation of state 
 $p=(\gamma-1)\rho$, the expansion is $a\sim t^{2/3\gamma}$
 , so $\gamma=2$ corresponds to the stiff fluid or 
 kinetic driven phase we have obtained $a\sim t^{1/3}$.

  Consider a universe created with Planck 
  radius ($\sim10^{-33}$ cm) and Planck
  density ($\sim 10^{93}$gcm$^{-3}$). If such a universe 
  expands to its present size greater than $\sim 10^{28}{\rm cm}$
then the density would be of order [11]
\begin{equation}
 \rho \sim 10^{93} \left (10^{28}/10^{-33} \right )^{-6}\;\simeq 10^{-273}
 {\rm gcm}^{-3}
 \end{equation}
  This should be compared to the present energy 
  density $\sim 10^{-30}{\rm gcm}^{-3}$ . Even if
  the kinetic energy density was immediately converted into dust the 
  resulting energy density would be $\sim 10^{-90}{\rm gcm}^{-3}$. 
  To account for this discrepancy
  we require the constant $A$ to be so large that the energy density is
  vastly greater than the Planck value for when 
  the universe is $\sim$ Planck size or equivalently
  the size of the universe is already much bigger than Planck 
  size for time $\sim$ Planck time
  ($t_{p}$). This problem we have outlined  
  we will call the {\em Planck density
  problem} [11] which is in many ways the most fundamental problem we first 
  need to try and solve with a cosmological model. It is present in 
  flat $k=0$ universe as the natural size of a radiation dominated universe
  with scale factor $a\sim t^{1/2}$, 
  with todays lifetime $10^{60}t_p$ is only $\sim 10^{-33}{\rm cm}*10^{30}
  \sim 10^{-3}{\rm cm}$! 

  The closely related {\em flatness problem} 
  is solved by having an exceedingly
  big value of $\dot{a}$ at the Planck time. This sets the density 
  parameter $\Omega$, where
  \begin{equation}
  \Omega=1+\frac{k}{\dot{a}^2}\;\;,
  \end{equation}
  extremely close to unity so that even today at time $\sim 10^{60}t_p$ 
  it still has not
  departed significantly from unity. Again the large 
  value of the constant $A$ can achieve
  this. 

  So far we have not included any inflationary early stage: indeed if 
  inflation 
  was actually required the big-bang model would have been rejected long ago!
  However, with inflation the  Planck density  problem 
  is helped by having 
  a huge expansion while the energy density remains roughly 
  constant. This  obviates the
  need for arbitrary constants that usually 
  set parameters, particularly $\dot {a}$, 
  vastly post-Planckian where quantum gravity 
  is utterly dominant. With inflation, the constant ``$A$'' is 
  automatically forced large  
  without requiring unnatural initial values -see eg.[18].

  There is a further problem of having a kinetic, or equivalently 
  stiff equation of state $\gamma=2$, driven early stage. There is 
  over production of gravitational waves in the resulting 
  universe [12]. 
  This was taken to rule out this equation of state and and more
  recent studies show eg.[13] that the scalar density fluctuation spectrum 
  is ``blue shifted''- too much power on small scales for galaxy formation.

  {\bf 3 Jordan or string frame description}\\
The field equations from the Brans-Dicke 
action (2) are well established and have
solutions [1]
\begin{equation}
a\propto
|t|^q\;\;\;;\;\;\Phi= |t|^{1-3q}
\end{equation}
where,
\begin{equation}
q=\frac{1+\omega\mp\sqrt{1+2\omega/3} }{4+3\omega}
\end{equation}
There are two different sets of solutions separated by a curvature
singularity at time $t_0$ which without loss of generality can be taken
to occur at time $t_0=0$. Note that there is an arbitrary constant that
can be introduced but we initially assume it to be unity. 

For the string theory case with $\omega=-1$ these solutions simplify for
times $t<0$
\begin{equation}
a\propto (-t)^{\mp\frac{1}{\sqrt{3}}}\;\;;\;\;\Phi= (-t)^{1\pm\sqrt{3}}
\end{equation}

The upper signs correspond to the inflationary behaviour while the lower
signs correspond to collapse -see Fig.(1). 
Incidentally this inflationary solution seems
$\em {fragile}$ in that the presence 
of any additional matter fields would tend
to destabilize it towards the collapsing solution. But we ignore this 
weakness and proceed with the pure vacuum dilaton model. The post-big bang 
solutions are simply got from equation 
(13) by substitution of $t\rightarrow -t$.
The required pre-big bang solution is
\begin{equation}
a\sim (-t)^{-\frac{1}{\sqrt{3}}}\;\;\;;\;\;\; \Phi \sim (-t)^{1+\sqrt{3}}
\;\Rightarrow l_p=(-t)^{-(1+\sqrt{3})/2}\;\;,
\end{equation}
while the expanding  post-big bang solution is
\begin{equation}
a\sim (t)^{\frac{1}{\sqrt{3}}}\;\;\;;\;\;\; \Phi\sim (t)^{1-\sqrt{3}}\;\;.
\end{equation}
This post-big bang phase is chosen at it can most easily be joined 
to a conventional FRW
expansion $a\sim t^{2/3\gamma}$, particulary radiation $a\sim t^{1/2}$. 
There is also a collapsing post-big bang branch which is presently ignored.
However the  field $\Phi$ has to fixed in the post-big bang phase
as there are strong constraints eg.[14] on any 
time evolution of $G\equiv \Phi^{-1}$ or equivalently the Planck length 
$l_p\equiv G^{1/2}$. Note that initially at $t\rightarrow
-\infty$, $ G\rightarrow 0$ and it grows towards infinity as $t\rightarrow
0_-$, this incidentally is the cause of the fragility as any matter will 
eventually dominate and cause collapse. 

In Fig.(1) we show the initial expanding pre-big bang solution 
undergoing a branch change so that it can match to a radiation dominated
phase. We assume that a branch change occurs when $G$ takes its present value
and remains constant in the subsequent post-big bang phase. 
Although this is somewhat ad-hoc ``quenching '' of the model it would seem to
give it the best chance of working 
as the ``wanted'' post-big bang solution  
unrealistically sets
$G=0$ again for $t=0_+$. The other collapsing post-big bang 
solution would have set $G=\infty$ 
at $t=0_+$.
This does rather constrain how the branch change 
should occur but there appear deficiencies in the model 
while it is still in the ``weak coupling'' domain. 

 Note that the scale factor and the Planck length 
 do not have the same
 time dependance $a\sim (-t)^{-0.6}$ 
 while $l_p\sim (-t)^{-1.4}$ -see Fig.(2). 
 At early times the scale factor starts larger 
 than the Planck length $a/l_p\sim (-t)$. Even so  
 the scale factor is infinitesimally
 small while the Planck length starts  even smaller and grows for 
 increasing time: we leave aside how it is consistent for strings to be 
 actually present. 
 As  $t\rightarrow 0_{-}$ the Planck length starts catching up with the 
 scale factor $a$ and overtakes it for $|t|<1$ - see Fig.(2). 
 This is not what one wants with inflation because we 
 are going to require $a>>l_p$ at the Planck time in the post-big bang phase
 in order to set the constant ``$A$'' there large and so avoid any Planck 
 problem. 
 
 Kinetic inflation is actually detrimental and is  
 taking the 
 initial state towards a quantum gravitational region and 
 away from what we require. As the time approaches the 
 singularity the Planck length together with  $G$ 
 increases until it presumably reaches its present value. Recall these 
 values then remain constant 
 during and after the branch change. But at this branch change 
 it is unclear why
 the scale factor is still much larger than $l_p$ 
 unless we again introduce an arbitrary  constant $A^{\natural}$ 
 on the pre-big bang side,
 where now $a=A^{\natural} (-t)^{-1/\sqrt{3}}$. 
 Otherwise to keep $a>>l_p$ the branch change has to occur long before
 the singularity is reached measured in the Planck time units of the
 post-big bang phase. This is still the  ubiquitous  
 mismatch of scales of the usual big bang model and entirely 
 analogous to choosing
 the constant $A$ in the massless scalar field model.  We suspect this 
 is true in all the alternative gravity theories, and certainly  
 Brans-Dicke with $\omega>-3/2$, 
 that display kinetic
 inflation as they all are conformally equivalent to a massless scalar field
 [8,10].

 The pre-big bang model is just equivalent to the massless scalar field, 
 no better or worse
 : it does not give an extra 
  mechanism for resolving the Planck 
  density problem. Further, the initial state at $t\rightarrow -\infty $ 
  seems rather contrived  
  in that while not 
 within the ``quantum foam'' it is still infinitesimally small. 
 The kinetic 
 inflation takes this state and drives it towards a reversed state where
 the Planck length is bigger than the scale factor: quantum gravitational 
 dominance. Before this can happen some other scale is intervening 
 and causing the natural $a\sim l_p$ scale at the 
 branch change to be broken to $a>>l_p$. One might 
 try to make the branch change account for this or have other
 inflationary mechanisms, but this would 
 negate any advantage the pre-big bang phase might have over other models.

 It might be objected that the scale factor 
 has no intrinsic meaning as a FRW
 metric is conformally invariant with 
 scale invariance $a\rightarrow \lambda a$. But our actual universe does not
 display this symmetry simply once masses are introduced
 : eg. Planck mass in our case. Our fixing of $\Phi$ in the  
 post-big bang phase breaks the underlying duality symmetries, but these 
 symmetries eg. $a\rightarrow 1/a$,
 are now obviously broken in the present universe. It is now crucial 
 that this actual value of the scale factor is ${\em large}$. 
 
 Another length scale is introduced when
 curvature is present and results in the flatness problem of why
 this length scale is so large. Usually this is resolved in a similar 
 fashion to the  Planck density problem but that cannot be assumed here. 
 Indeed it has already been shown that the pre-big bang model 
 is vulnerable to curvature dominating [15]. We can see this roughly  since
 $\dot{a}\sim (-t)^{-(1/\sqrt{3})-1}$ tends to zero as $t\rightarrow -\infty$
 cf. eq.(10), and 
 any curvature present will dominate (recall $G\rightarrow 0$ in this domain
 so that LHS of Einstein's equations are isolated).
 Depending on whether $k=\pm1$ the scale factor will either be infinite 
 or contract to a singularity as $t\rightarrow -\infty$ [15]. 
 Either way other
 mechanisms would be required to explain these initial conditions and which
 relegates the question of whether there is subsequent inflation. We note in
 passing that the initial conditions based on a canonical classical measure
 do suggest an infinite scale factor [16]- although there are a 
 number of caveats with such arguments[17].

{\bf 4 Quantum fluctuations as $t\rightarrow 0_-$}\\
In scalar potential driven inflationary models there is the possibility of
the fluctuations being so large that eternal inflation results, where many
separate universes are created. In the potential driven case this is 
possible for energy densities below Planck values[18]. While trying to find 
the analogous thing for kinetic driven inflation we were first 
led to doubts about this  type of inflation.
In kinetic driven inflation the only scale for fluctuations to dominate 
is now for Hubble parameter $H=(-t)^{-1}\geq1$ 
but the scale factor is only $\sim 1$ at this time. Already
by this stage the  quantum fluctuations are dominating over the classical  
solution and easily cause a shift to the collapsing solution [19].
This is consistent with our criticism that the universe is not adequately
being inflated to be large and classical as it can so easily be totally 
switched to the collapsing phase.         

The fluctuation spectrum for scalar density 
perturbations has the same worrisome ``blue
spectrum'' as that of a massless scalar field-see eg.[13]. 
It also has gravitational waves [20] 
which earlier were taken to rule out the massless scalar field or stiff 
equation of state model [12]. 
It seems ironic that this is being made a virtue,
or at least a major point of interest[1].

 {\bf 5 Quantum cosmology and branch changes}\\ 
 It has been suggested that quantum cosmology could explain the branch change
 in this or related models that might avoid the problems we have 
 outlined [5]. This is very suspect  
 as a quantum description should not be 
 relevant after an inflationary phase when the universe is driven large 
 and classical.  It is certainly 
 not tidy to require quantum cosmology twice, once for
 the initial condition at $t\rightarrow -\infty$ and then also to avoid a
 singularity at time $t=0$. In general, quantum gravitational 
 phenomena should only occur  
 over small volume regions and then they would be expected to cause 
 large perturbations that probably need an inflationary 
 mechanism to remove them. 
 Indeed, quantum cosmology is a source of the  Planck density problem,
 it requires inflation to expand the initial $\sim$ Planck size region it
 usually predicts see eg.[18].\\ 
 {\bf 6 Outlook and conclusions}\\
 What can be learnt from this model? Most of the misunderstanding 
 seems to have come from taking inflation to be defined as 
 a change in the comoving
 Hubble length $(Ha)^{-1}\equiv \dot{a}^{-1}$, which suggests near 
 infinite expansion of $a$ is possible. This has not picked up 
 the inadequacy of kinetic inflation for solving 
 the Planck density problem.
 Indeed using this definition has led one to conclude that the contracting
 phase of a 
 massless scalar field in the Einstein frame is an adequate 
 inflationary model [21]-
 an obvious inadequacy of any realistic model. Most 
 previous work does not seem
 to have adequately made 
 contact with our actual universe which  also no longer  displays duality
 symmetries, but rather has a fixed value of Newton's constant. An 
 exception is ref.[22] where similar doubts have been 
 expressed about reconciling the pre-big bang scenario with the actual
 universe.

   What about using the contracting phase in these model, such a phase 
   does display a mechanism for describing fluctuations as
they are ``left outside'' the contracting Hubble radius [1]. 
Such a contraction
might be incorporated into a cosmological model 
along the lines of refs.[23,24]. Of course such contractions have nothing 
per se to do with inflation, recall pure radiation Fig.(1). 
But they do 
beg the question of why the universe is large in the first place and how
does one avoid the inevitable 
approaching singularity? One needs a ``bounce'' 
that requires violation of the 
strong-energy condition [23,24]. In ref.[24] one uses
the two post-big bang solutions and creates a bounce between them, 
although this uses 
other coupled matter fields to violate the strong-energy condition. If
one is anyway going to violate the strong-energy condition one might as well
use inflation as this can also explains the Planck problem together with a
mechanism for creating fluctuations see eg. [18]. 

In conclusion, the pre-big bang model is only equivalent to an ordinary
(strong-energy satisfying) big bang model.  
If they are to explain our present universe, both suffer, from requiring
constants that exceed natural Planck values drastically. This is 
unlike potential driven inflation, although we leave aside arguments 
that that itself suffers from other  problems of fine tuning. 

The kinetic driven inflation is defective as the Planck length
grows faster than the scale factor so sending any universe towards
quantum gravitational dominance. kinetic inflation is a 
mirage: it goes away in
the Einstein frame, and cannot inflate to give our {\em large} universe.\\

{\bf Acknowledgement}\\
I should like to thank Janna Levin, David Wands and Andrei Linde 
for helpful discussions.\\
{\large Figure Captions}\\
{\bf Figure 1)}.\\ 
The pre-big bang scenario. The two solutions to eq.(11) are plotted 
for $\omega=-1$ -solid
lines. The required sequence of solutions are indicated by the
solid arrows. The wanted expanding 
solution $a\sim (-t)^{-1/\sqrt{3}}$ starting at time $t=-\infty$ undergoes a
branch change when $t\simeq 0$ to the 
expanding post-big bang solution. It can then
connect to the radiation dominated universe (dotted line). Note that the
two unwanted collapsing branches (open arrows) of eq.(11) 
are conveniently ignored in this scenario.
Pure radiation has a collapsing phase for negative time.
\\
{\bf Figure 2)}.\\
The Planck problem of the pre-big bang model. The scale factor (solid 
lines) 
$a=(-t)^{-1/\sqrt{3}} $for ($t<0$); 
and that of radiation $a\sim t^{1/2}$ for $(t>0)$. 
The Planck length (dotted lines) during the pre-big bang phase grows more
rapidly than scale factor: in the 
post-big bang region it is fixed to its present
value. Without introducing  arbitrary large constants  it is not clear why
$a>>l_p$ at $t\simeq t_p=1$ after the branch change occurs
. The resulting universe
suffers a Planck problem identical to that of conventional big bang model
without inflation being present.\\

\newpage

{\bf References}\\
\begin{enumerate}
\item M. Gasperini and G. Veneziano, Astropart. Phys. 1 (1993) p.317.\\
M. Gasperini,``Birth of the Universe in String cosmology'', preprint 
gr-qc/9706037.\\
S.J. Rey, ``Recent progress in string inflationary cosmology'',
preprint hep-th/9609115\\
J.J. Levin, ``Gravity driven inflation'', preprint gr-qc/9506017.\\
Available at pre-big bang web site:\\ 
http://www.to.infn.it/teorici/gasperini/
\item M.D. Pollock and D. Shahdev, Phys. Lett. B 222 (1989) p.12.
\item R. Brustein and G. Veneziano, Phys. Lett. B 329 (1994) p. 429.
\item  S.J. Rey, Phys. Rev. Lett. 77 91996) p.1929.
\item M. Gasperini, J. Maharana and G. Veneziano, Nucl. Phys. B 472 (1996)
p.349.\\
M. Gasperini and G. Veneziano, Gen. Rel. Grav. 28 (1996) p. 1301.\\
J.E. Lidsey, Phys. Rev. D 55 (1997) p.3303.
\item E.S. Fradkin and A.A. Tseytlin, Phys. Lett. B 158 (1985) p.316.\\
C.G. Callan, D. Friedan, E.J. Martinec and M.J. Perry, Nucl. Phys. B 262
(1985) p.593.\\
C. Lovelace, Nucl. Phys. B 273 (1985) p.135.
\item G. Veneziano, Phys. Lett. B 263 (1991) p. 287.
\item S. Kalara, N. Kaloper and K.A. Olive, Nucl. Phys. B 341 (1990) p.252.
\item R. Brustein and P.J. Steinhardt, Phys. Lett. B 302 (1993) p.196.
\item G. Magnano and L.M. Sokolowski, Phys. Rev. D 50 (1994) p. 5039.\\
L.M. Sokolowski, preprint gr-qc/9511073.
\item Y.B. Zeldovich, ``My Universe: selected reviews'', Harwood Academic
Press (1992)
\item Y.B. Zeldovich and I.D. Novikov, ``The structure and evolution of
the universe: relativistic astrophysics vol. 2'', Chicago University Press
(1983) p.666.
\item J. Hwang and H. Noh, Phys. Rev. D 54 (1996) p. 1460.\\ 
``Density spectrums from kinetic inflations'', preprint gr-qc/9612065
\item J.D. Barrow and P. Parsons, Phys. Rev. D 55 (1997) p.1906.
\item M.S. Turner and E.J. Weinberg, Phys. Rev. D 56 (1997) p.4604. 
\item S.W. Hawking and D.N. Page, Nucl. Phys. B 298 (1988) p. 789.
\item D.H. Coule, Class. Quant. Grav. 12 (1995) p.455.
\item A.D. Linde, ``Particle Physics and Inflationary cosmology'', 
Harwood press: Switzerland (1990).
\item Z. Lalak and R. Poppe,``Scalar field fluctuations in pre-big bang 
cosmologies'', preprint gr-qc/9704083.
\item J. Hwang, ``Gravitational wave spectrums from kinetic inflations'',
preprint gr-qc/9710061.
\item M. Gasperini and G. Veneziano, Mod. Phys. Lett. A 8 (1993) p. 3701.
\item J.J. Levin, Phys. Rev. D 51 (1995) p. 462. \\
{\em ibid} p.1536.
\item R. Durrer and J. Laukenmann, Class. Quant. Grav. 13 (1996) p. 1069.\\
\item F.G. Alvarenga and J.C. Fabris, Class. Quant. Grav. 12 (1995) p. L69.
\end{enumerate}
\end{document}